\documentclass[aps,prb,twocolumn,groupaddress]{revtex4}

\usepackage{graphicx}
\usepackage[hypertex]{hyperref}
\usepackage{amsmath,amssymb}
\usepackage{bm}

\newcommand{\kv}{\mathbf{k}}

\newcommand{\qv}{\mathbf{q}}
\newcommand{\rv}{\mathbf{r}}

\newcommand{\yv}{\mathbf{y}}

\newcommand{\Bv}{\mathbf{B}}

\newcommand{\Kv}{\mathbf{K}}

\newcommand{\al}{\alpha}        %We define shorthands for greek symbols%

\newcommand{\ga}{\gamma}

\newcommand{\de}{\delta}
\newcommand{\De}{\Delta}
\newcommand{\ep}{\epsilon}

\newcommand{\lam}{\lambda}

\newcommand{\si}{\sigma}
\newcommand{\Si}{\Sigma}

\newcommand{\om}{\omega}

\newcommand{\Tr}{\text{Tr}}

\newcommand{\scr}{\text{scr}}

\newcommand{\eff}{\text{eff}}

\newcommand{\QCP}{\text{QCP}}

\newcommand{\avv}{\text{av}}
\newcommand{\sep}{\text{sep}}
\newcommand{\reff}{\text{ref}}

\newcommand{\eV}{\text{eV}}  %We define text as by \text%

\newcommand{\meV}{\text{meV}}
\newcommand{\nm}{\text{nm}}

\newcommand{\hb}{\hbar}

\newcommand{\la}{\langle}
\newcommand{\ra}{\rangle}
\newcommand{\da}{\dagger}

\newcommand{\ti}{\tilde}

\newcommand{\qaq}{\quad \text{and} \quad}       %We define \quad .. \qaud combinations%

\newcommand{\bpm}{\begin{pmatrix}}              %We define the envirioment shorthands%
\newcommand{\epm}{\end{pmatrix}}
\newcommand{\bi}{\begin{itemize}}
\newcommand{\ei}{\end{itemize}}
\newcommand{\be}{\begin{equation}}
\newcommand{\ee}{\end{equation}}
\newcommand{\bml}{\begin{multline}}
\newcommand{\eml}{\end{multline}}

\begin{document}

\title{The Influence of Remote Bands on Exciton Condensation in Double-Layer Graphene}

\author{M. P. Mink}
\email{m.p.mink@uu.nl}
\affiliation{Institute for Theoretical Physics, Utrecht
University, Leuvenlaan 4, 3584 CE Utrecht, The Netherlands}

\author{A. H. MacDonald}
\affiliation{Department of Physics, The University of Texas at Austin, Austin Texas 78712}

\author{H. T. C. Stoof}
\affiliation{Institute for Theoretical Physics, Utrecht
University, Leuvenlaan 4, 3584 CE Utrecht, The Netherlands}

\author{R. A. Duine}
\affiliation{Institute for Theoretical Physics, Utrecht
University, Leuvenlaan 4, 3584 CE Utrecht, The Netherlands}
\date{\today}

\begin{abstract}
We examine the influence of remote bands on the tendency toward exciton condensation in a system consisting of two parallel graphene layers with negligible interlayer tunneling. We find that the remote bands can play a crucial supporting role, especially at low carrier densities, and comment on some challenges that arise in attempting quantitative estimates of condensation temperatures.
\end{abstract}

\maketitle
\vskip2pc

\section{Introduction}
In the last few years, there has been enormous experimental and theoretical interest in the properties of graphene, a hexagonally ordered two-dimensional sheet of carbon atoms.\cite{neto} Graphene is a gapless semiconductor whose valence and conduction bands cross linearly
near two inequivalent Dirac points (valleys) located at the honeycomb lattice Brillouin zone corners.
Since the Fermi level in graphene can be shifted by the electric field effect,\cite{novoselov}
the density and polarity of the charge carriers can be tuned by application of a gate voltage.

A symmetric voltage applied between the layers of a double-layer graphene system can induce an electron density in one layer, and an equal hole density in the other.
Under these circumstances electrons will tend\cite{lozovik,zhang,min} to form a broken symmetry state in which coherence is established spontaneously between separate layers in the absence of interlayer tunneling.
The broken symmetry state is driven by Coulomb interactions between layers and favored by nesting between the nearly circular Fermi surfaces. In this state the one-particle electron density matrix has nonzero components that are off-diagonal in the layer indices. A bilayer spontaneous coherence state is a type of exciton condensate (see below) which has particularly interesting transport properties\cite{jpe,jungjung} when the two layers can be contacted independently.  In graphene the possibility of independently contacting two graphene sheets separated by a distance of several
nanometers has recently been demonstrated.\cite{schmidt,tutuc}
After performing a particle-hole transformation on the hole-like layer, the broken-symmetry state can be viewed as a
spatially indirect electron-hole pair (exciton) condensate.
This is the viewpoint we will use throughout this paper.

\begin{center}
\begin{figure}
\includegraphics[width = 0.47 \textwidth]{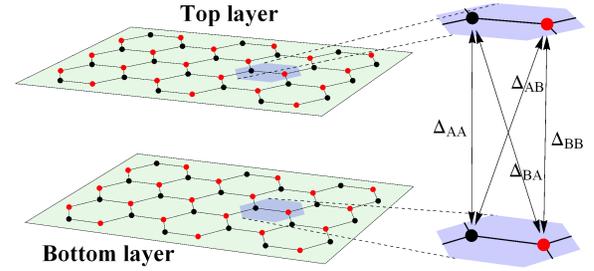}
\caption{\label{fig_dlg} (Color online) Double-layer graphene consists of two parallel graphene sheets. The black dots are atoms in sublattice $A$ and the red dots in sublattice $B$. The four component structure of the condensate is illustrated at right. The exact definition of the order parameter $\De$ is given in Sec.~\ref{sec_met}.}
\end{figure}
\end{center}

In addition to having spin and valley degrees of freedom,
electron states in single layer graphene are two-component spinors with a sublattice degree of freedom.
As a consequence, the order parameter for exciton condensation has a four-component sublattice structure.
The structure of the double-layer graphene system is illustrated in Fig.~\ref{fig_dlg}, where the blow-up at right emphasizes the 4-component structure of the order parameter.
In the broken-symmetry state all elements of the expectation value $\la c^\da_{t,\al} c_{b,\al'} \ra$, where $c^\da_{t,\al}$ is the creation operator of an
electron in the top layer in sublattice $\al = A,B$, and $c_{b,\al'}$ is the annihilation operator of an electron in the bottom layer in sublattice $\al' = A,B$,
can be nonzero at each crystal momentum.

This system was first studied by Lozovik and Sokolik \cite{lozovik} for $k_F d \gg 1$, with $k_F$ the Fermi momentum and $d$ the interlayer distance, and
later revisited by a number of authors. \cite{zhang,min,kharitonov,seradjeh,bistritzerar,kharitonov2} Min {\it et al.} \cite{min} evaluated the ground state superfluid density of the condensate,
demonstrating that it is proportional to the carrier Fermi energy, and on this basis
proposed that the Kosterlitz-Thouless (KT) transition temperature for exciton condensation could be of the order of room temperature in systems with high carrier densities and small layer separations. This suggestion was countered by Kharitonov {\it et al.}, \cite{kharitonov} who
estimated that $T_c < 1\text{ mK}$ because the interlayer interaction is strongly screened in the normal state.
In this paper we consider two graphene layers
without any dielectric so that the effective dielectric constant for interlayer interactions $\ep=1$.
Our work differs from earlier work in that we include the influence of remote bands, which play a supporting role.
When a static screening approximation is employed we find that the maximum critical temperature is
of the order of Kelvins and occurs at rather low carrier densities.
Since static screening underestimates the interlayer interaction, particularly for
remote band effects, higher condensation temperatures appear to be a possibility.

This paper is organized as follows. In Sec.~\ref{sec_res} we present the main results of our analysis for the condensate structure and the transition temperature $T_c$. In Sec.~\ref{sec_conc} we present our discussion and conclusions. Sec.~\ref{sec_met} describes important technical details of the calculations
used to obtain the results described in Sec.~\ref{sec_res}.

\section{Results}
\label{sec_res}
In this paper we estimate the transition temperature and determine the optimal structure for exciton condensation in double-layer graphene.
The details of this calculation are given in Sec.~\ref{sec_met}. We consider only order parameters that have zero center-of-mass momentum. Using mean-field theory we derive a self-consistent gap equation, which we solve numerically in two approximations. First, to gain physical insight, we approximate the screened Coulomb interaction by a contact interaction. We determine the strength of the contact interaction by performing an angular average over incoming and outgoing momenta on the Fermi surface.
In this version of our calculation, we use the full graphene dispersion, but we find that
substituting the Dirac dispersion with an appropriate ultra-violet cutoff gives nearly identical results.
In an effort to obtain more quantitatively reliable results, we have also used a separable
approximation to the screened Coulomb interaction,
similar to the form used by Lozovik {\it et al.}, \cite{lozovik2} to determine the gap at zero temperature.
This calculation is carried out within the Dirac approximation. The greatest source of uncertainty in our work is the approximation used for the screened interlayer interaction. In all of our explicit calculations we use a static approximation which overestimates screening. The critical temperatures we obtain will therefore tend to be underestimated.

In the present section we present the results of our calculations.
We describe the optimal condensate structure, which turns out to be the same in both approximations.
We then discuss the phase diagrams we obtain.
The contact-interaction approximation leads to unphysically high values for the transition temperature, as may be expected for such a crude model. In contrast, using the separable approximation for the screened Coulomb interaction, we obtain a $T_c$ of order of Kelvins for typical carrier Fermi energies $V_g=0.25-0.5\ \meV$ and interlayer distances $d<4\ \nm$, respectively.

\begin{center}
\begin{figure}
\includegraphics[width = 0.47 \textwidth]{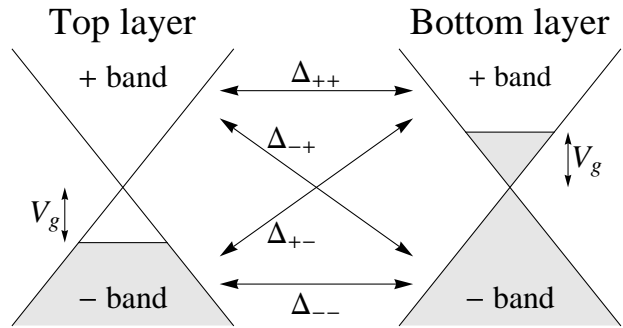}
\caption{\label{fig_Dband} Dirac approximation band dispersions for the top and bottom layer. The
shaded areas correspond to filled states. The four components of the order parameter
in the band representation are indicated. In the close-band approximation only the (lower energy) valence band in the top layer (bottom left) and the (higher energy) conduction band of the bottom layer (top right) are retained.}
\end{figure}
\end{center}

\subsection{Condensate structure}
In the condensed state, the order parameter sublattice structure ($\De_{\al,\al'}(\kv)$ with $\al,\al'= A,B$)
is optimized.  (The explicit definition of the order parameter $\De$ is given in Sec.~\ref{sec_met}.)
We find that the structure that is optimal has the form $\bm{\De} \equiv (\De_{AA},\De_{AB},\De_{BA},\De_{BB})= (\De_1,\De_2,-\De_2,-\De_1)$, where $\De_1 \gg \De_2$.
The property that $\De_{AA} = - \De_{BB}$ can be understood in terms of the primary mechanism by which condensate formation
lowers the energy of the system, namely the opening of an avoided crossing gap between the conduction band of one layer
and the valence band of the other layer.  In the Dirac-band approximation the conduction and valence band sublattice spinors are
$(1,\exp(i\phi_{\bf k}))/\sqrt{2}$ and $(1,-\exp(i\phi_{\bf k}))\sqrt{2}$ respectively, where ${\bf k}$ is momentum
measured from the Brillouin-zone corner and $\phi_{\bf k}$ is the angular orientation of this momentum.
An order parameter with $\De_{AA} = - \De_{BB}$ couples these spinors with equal strength at all $\phi_{\bf k}$.

We note that the same order parameter sublattice structure is associated with the broken inversion symmetry
often thought\cite{ChiralSymmetryBreaking} to be plausible in isolated graphene sheets.
In that case, of course, the mean-field potential couples sites in the same layer.
Broken inversion symmetry in an isolated graphene sheet is analogous to chiral symmetry breaking in
elementary particle physics and is most likely to occur in neutral sheets without any carriers.
Experiments appear to establish that spontaneous gaps do not in fact occur in single-layer graphene;
angle-resolved photoemission experiments\cite{ARPES} are perhaps most conclusive in this respect.
Spontaneous gaps do however appear\cite{Ong} in neutral graphene sheets when a magnetic field is applied.
It is sometimes argued\cite{Magnetic_Catalysis} that the appearance of gaps in a field demonstrates that
this order is barely avoided and latent even in the absence of a field.  (For a contrary view see Ref.~[{~\onlinecite{QHF}}]).
We will show that, because of their order parameter compatibility,
latent sublattice-pseudospin chiral symmetry breaking order in a single layer
is favorable for bilayer excitonic order.  Although the presence of carriers always acts against order in an
isolated graphene layer, we show that in the bilayer case it acts in favor of order provided
that the carrier densities have opposite signs in opposite layers.

\begin{center}
\begin{figure}
\includegraphics[width = 0.47 \textwidth]{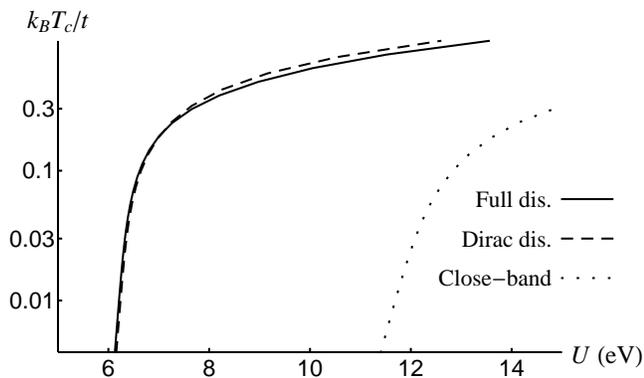}
\caption{\label{fig_pdUT1} $T_c(U)$ for the contact-interaction approximation for $V_g=0.1\ \eV$.
Here, $U$ is the effective interaction strength in {\rm eV} units. The critical temperature $T_c$ is
expressed in units of the hopping parameter $t= 2.8 \ \eV$.
The solid line is obtained using the full dispersion, the dashed line using the Dirac approximation,
and the dotted line using the close-band approximation.}
\end{figure}
\end{center}

\subsection{Full dispersion and close-band approximation}
In this section we consider the influence of the full graphene dispersion and the close-band approximation on our result for the transition temperature. Since we are interested in qualitative effects, it is sufficient to consider the contact-interaction approximation results. We show in Fig.~\ref{fig_pdUT1} the critical temperature $T_c$ as a function of the effective interaction strength $U$ for a fixed carrier Fermi energy $V_g = 0.1\ \eV$. The transition temperature $T_c$ is scaled by the hopping strength $t = 2.8\ \eV$. Note that the interaction strength $U$ in our model has the units of energy and that it is obtained as the continuum model interaction strength divided by the unit cell area. The solid line is obtained using the full graphene dispersion and the dashed line using the Dirac approximation. When we employ the close-band approximation, we obtain the dotted curve. Below we relate the parameter $U$ to the contact-interaction approximation to the screened Coulomb interaction: $V^{\scr}(\qv) \to U$, where $V^{\scr}(\qv)$ is the screened interlayer interaction matrix element.

When we want to employ the Dirac approximation, we are confronted with the fact that the theory has one free parameter, namely the value of the cutoff energy $\xi$ that is needed to cure the ultra-violet divergence in the gap equation
on which $T_c$ will depend. In Sec.~\ref{sec_met} we explain how to choose $\xi$, so that the results obtained using the Dirac dispersion and the full graphene dispersion overlap for temperatures corresponding to the low energies where the two dispersions do not differ noticeably. With this choice of $\xi$, we obtain the dashed curve in Fig.~\ref{fig_pdUT1} for the Dirac model and the solid curve for the full $\pi$-band model. Note that the curves differ only slightly and then only for temperatures $k_B T > 0.2 t$ at which the difference between the full graphene dispersion and Dirac dispersion becomes noticeable. Since the results obtained using the Dirac and full dispersions do not differ appreciably, we use the Dirac approximation for the contact-interaction phase diagrams shown in the next section.  The dotted curve in Fig.~\ref{fig_pdUT1} is the result of the close-band approximation, discussed further below.

Notice that the critical temperature is weakly dependent on
interaction strength at large $U$ and strongly-dependent on interaction strength at small $U$.
The weak dependence occurs in the range of values where $U$ is strong enough to produce order
even when the carrier Fermi energy is set to zero, and the strong temperature-dependence
in the range of $U$ where order is assisted by the gate-driven Fermi surface nesting.
To understand this observation consider the linearized $T_c$ equation which is derived in detail in Sec.~\ref{sec_met}
and represented there by Eq.~(\ref{eq_Tc1D}):
$$
\frac{1}{U} =\frac{1}{2} \int_{0}^\xi d \ep \nu(\ep) \sum_{s=\pm 1} \frac{1-2 n_f(\ep + s V_g) }{\ep + s V_g}.
$$
Here, $\nu(\ep)$ is the isolated layer density-of-states per spin and valley per unit cell,
$n_f(\ep)$ is  the Fermi distribution function, and $s=-$ and $s=+$ terms
correspond respectively to close and remote band contributions.  (The cross
contributions between these two pairs vanish identically in the linearized gap equation.)
When the carrier Fermi energy $V_{g} \to 0$ the two types of particle-hole pairs make identical contributions to the
right-hand-side that decrease with temperature on the scale of the band width $\sim 3t$ and
approach $1/2t $ for $T \to 0$.  (The $1/t$ dependence for $T \to 0$ can
be understood by noting that $\nu(\ep) \propto \epsilon/t^2$.)  The linearized gap equation
for $V_{g} \to 0$ has a solution only if $U/t \gtrsim 2$.   For $V_{g} > 0$ the situation changes.
The remote band contribution decreases slightly as these bands are separated from the
Fermi energy.  At the same time the close band contribution from small $\epsilon$ is
enhanced, and in fact diverges logarithmically as $T \to 0$ because the density of states $\nu(V_g)$ is finite when the energy denominator vanishes.  This is the Fermi surface nesting effect.
Because the right-hand-side diverges, the linearized gap equation will always have a
solution.  On the other hand because the divergence is only logarithmic, the
critical temperature will be extremely small if $U$ is well below the strength at
which coherence already appears for $V_{g} \to 0$.  This observation accounts for the
rapid drop in $T_c$ at a particular interaction strength.  Because the
high-energy contributions of close and remote bands are nearly identical, the interaction strength
required for a high critical temperature is badly underestimated if the remote bands are ignored.
Note that the lowest temperature illustrated in Fig.~\ref{fig_pdUT1} $\sim 0.003 t$ is $\sim 100 {\rm K}$.

\subsection{Phase diagrams}
\begin{center}
\begin{figure}
\includegraphics[width = 0.45 \textwidth]{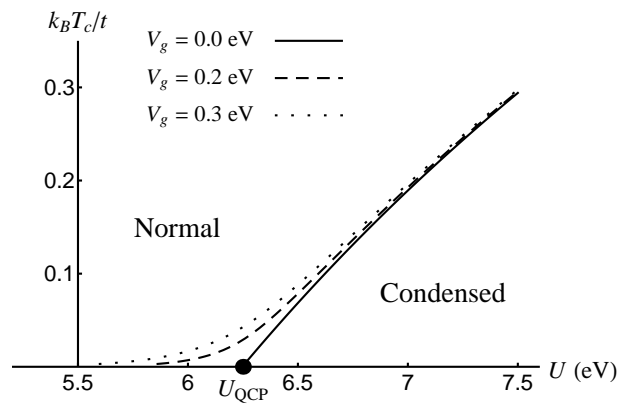}
\caption{\label{fig_pdUT2} The scaled critical temperature $T_c/t$ versus the effective interaction strength $U$ in eV's for several values of the carrier Fermi energy $V_g$ for the contact interaction approximation. From right to left the curves correspond to $V_g = (0, 0.2, 0.3)\ \eV$. The location of the normal state and condensed state are indicated. The quantum critical point is marked by a dot.}
\end{figure}
\end{center}

In this section we discuss the phase diagrams constructed from contact-interaction and
separable potential approximations. We start with the former to gain physical insight and then compare with the latter.
In Fig.~\ref{fig_pdUT2}, we show how the contact-interaction transition temperature versus effective interaction strength curve depends on carrier density.  From right to left the three curves correspond to the carrier Fermi energies
$V_g = (0, 0.2, 0.3)\ \eV$. In Fig.~\ref{fig_pdUT2} we sees that $T_c$ is an increasing function of both $U$ and, for a fixed value of $U$, also an increasing function of $V_g$.\\
This figure provides a nice illustration of the separate roles played by the carriers (the conduction band
states occupied by electrons and the valence band states occupied by holes) and
higher energy states in forming the instability. The contribution of the carriers
to the linearized gap equation scales with the density-of-states at the Fermi level, and hence with $V_{g}$.
The higher energy contribution depends on the density-of-states far away from the Dirac point near the
model's cut-off energy.
For $V_g=0$ only the remote band contribution is present and the system has a quantum critical point at
$U_{\QCP} = 6.25\ \eV$ which is indicated in Fig.~\ref{fig_pdUT2} by the dot.
Because of the $1/E$ weighting factor in the gap equation combined with the linear in $E$ density-of-states of the
Dirac model, this quantum critical point satisfies a Stoner-like criterion $\nu(\xi) U=1$, where $\nu(\xi)$ is the density of states at the cutoff energy $\xi$.
When $V_g \ne 0$, condensation occurs at any value of $U$,
but not at low temperatures unless $U$ is close to $U_{\QCP}$.
The carrier contribution has a larger relative importance, and values of $U$ that are
substantially smaller than $U_{\QCP}$ can still yield high transition temperatures
when $V_{g}$ is pushed toward the largest physically realistic values $\sim 0.3\ {\rm eV}$.
The high energy state contribution to the gap equation can be captured approximately by replacing the interaction among carriers by the effective interaction $U_{\eff} = U/(1-U/U_{\QCP})$. We note that room temperature condensation corresponds to $k_{B} T_{c} \simeq 10^{-2} t$, which is a low-temperature on the
scale of this phase diagram.

\begin{center}
\begin{figure}
\includegraphics[width = 0.45 \textwidth]{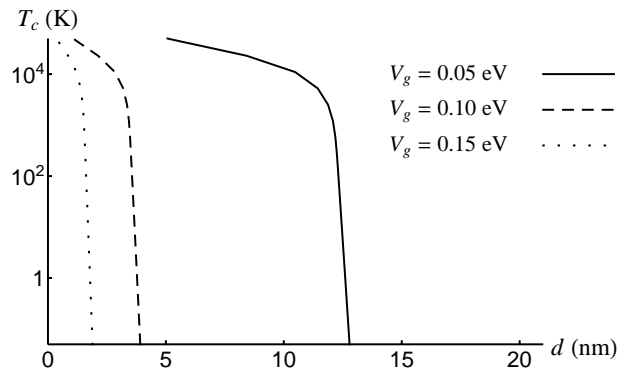}
\caption{\label{fig_pddT} $T_c(d)$ critical curves calculated in the Dirac approximation with the transition temperature $T_c$ in Kelvin versus the interlayer distance $d$ in nanometer for the contact-interaction approximation. From right to left the curves correspond to $V_g = (0.05, 0.1, 0.15)\ \eV$. To the right of or above a curve the system is in the normal phase, to the left of or below a curve the system is in the condensed state.
}
\end{figure}
\end{center}

The $T_c$ prediction based on the constant interaction model depends on a procedure for constructing an accurate
value of $U$.  We discuss such a procedure in Sec.~\ref{sec_met} where we derive an expression
in which $U$ depends on Fermi energy $V_g$ and interlayer distance $d$. We estimate $U$ by performing the angular average of the screened interlayer Coulomb interaction over incoming and outgoing momenta restricted to the Fermi surface. The phase diagram thus obtained is shown in Fig.~\ref{fig_pddT}, where we plot the transition temperature $T_c$ in Kelvin versus the interlayer distance $d$ in nanometers.  From right to left the curves correspond to carrier Fermi energies $V_g = (0.05, 0.1, 0.15)\ \eV$. We observe that $T_c$ is a decreasing function of $d$, as expected. The behavior of $T_c$,
which decreases as a function of $V_g$, is opposite to the behavior observed in Fig.~\ref{fig_pdUT2} in which $V_{g}$ was varied at fixed $U$.  This behavior illustrated in Fig.~\ref{fig_pddT} occurs because screening increases
with density-of-states and hence with $V_g$, decreasing the value of $U$.  Both the high transition temperatures and
the trends illustrated in this figure are suspect, however, because static screening at the Fermi energy is irrelevant for the high energy states in the linearized gap equation.  This conundrum demonstrates that reliable $T_c$ estimates
are challenging.

\begin{center}
\begin{figure}
\includegraphics[width = 0.47 \textwidth]{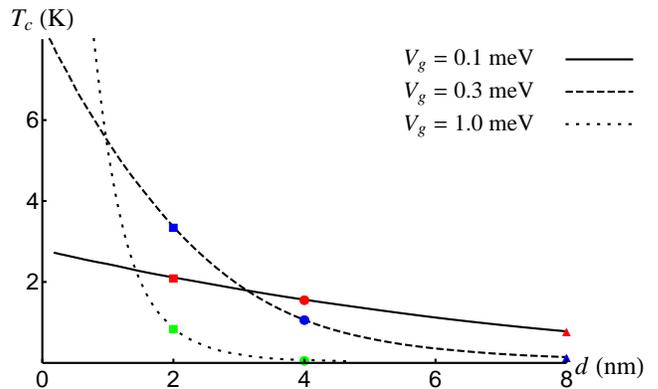}
\caption{\label{fig_pd2Td} (Color online) Transition temperature $T_c$ in Kelvin versus interlayer distance $d$ in nanometer for three fixed carrier Fermi energies for the separable-potential approximation. The solid line corresponds to $V_g = 0.1\ \meV$, the dashed to $V_g = 0.3\ \meV$, and the dotted to $V_g = 1\ \meV$. The points corresponding with Fig.~\ref{fig_pd2TV} are indicated by the colored markers.
}
\end{figure}
\end{center}

\begin{center}
\begin{figure}
\includegraphics[width = 0.47 \textwidth]{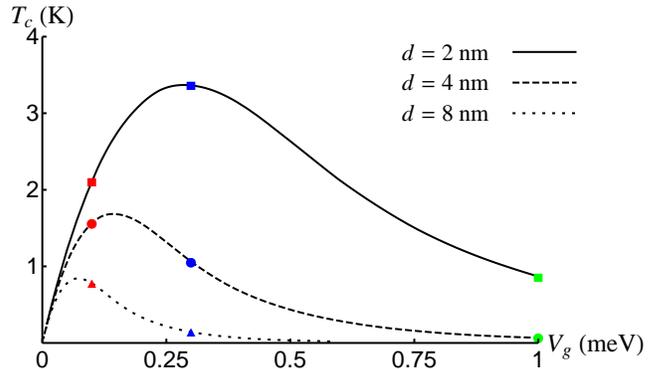}
\caption{\label{fig_pd2TV} (Color online) The critical lines of transition temperature $T_c$ in Kelvin versus carrier Fermi energy $V_g$ in milli-electronvolt for three fixed interlayer distances for the separable-potential approximation. The solid line corresponds to $d = 2\ \nm$, the dashed to $d = 4\ \nm$, and the dotted to $V_g = 8\ \nm$. The points corresponding with Fig.~\ref{fig_pd2Td} are indicated by the colored markers.
}
\end{figure}
\end{center}

In Sec.~\ref{sec_met} we show how to obtain the transition temperature $T_c$ as a function of interlayer distance $d$ and carrier Fermi energy $V_g$ using a separable-potential approximation. In Fig.~\ref{fig_pd2Td} we plot the resulting transition temperature $T_c$ in Kelvin versus interlayer distance $d$ in nanometer for three fixed carrier Fermi energies. The solid line corresponds to $V_g = 0.1\ \meV$, the dashed to $V_g = 0.3\ \meV$, and the dotted to $V_g = 1\ \meV$.  The separable
potential approximation is more realistic and at small $d$ captures the expected increase of $T_c$ with $V_g$.
In Fig.~\ref{fig_pd2TV} we plot the separable potential transition temperature $T_c$ in Kelvin versus carrier
Fermi energy $V_g$ in meV for three fixed interlayer distances. The solid line corresponds to $d = 2\ \nm$, the dashed to $d = 4\ \nm$, and the dotted to $d = 8\ \nm$. Corresponding points are indicated by colored markers in the two Figs.~\ref{fig_pd2Td} and \ref{fig_pd2TV}. In Figs.~\ref{fig_pd2TV} we see that $T_c$ has a non-monotonic dependence on $V_g$. This
behavior is due to a competition between the increase of screening (as in Fig.~\ref{fig_pddT}) and the increase of the Fermi energy for larger $V_g$ (as in Fig.~\ref{fig_pdUT2}). The maximum $T_c$ for fixed $d$ occurs for
$k_F d \simeq 0.001$ and the height of the maximum goes as $1/d$.  This conclusion is, however, also based on
an approximation scheme that is unreliable for the high energy virtual states that appear in the gap equation. We note that in obtaining Figs.~\ref{fig_pd2Td},\ref{fig_pd2TV} we approximated the polarizibality of Graphene by a constant, neglecting the increase of $\Pi$ with scattering momenta $q>2 k_F$. Including this dependence will lead to a suppression of the transition temperatures shown in Figs.~\ref{fig_pd2Td},\ref{fig_pd2TV}.

\begin{center}
\begin{figure}
\includegraphics[width = 0.47 \textwidth]{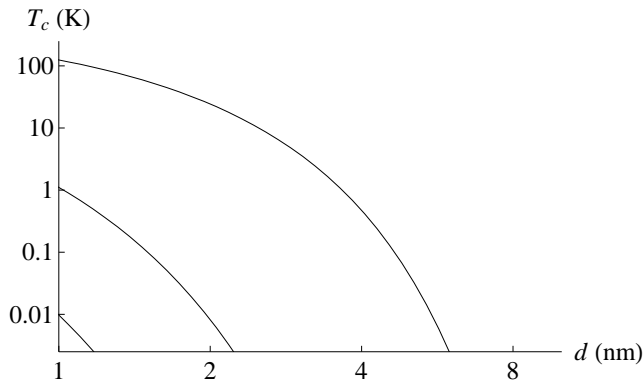}
\caption{\label{fig_phenpic} The critical lines of transition temperature $T_c$ in Kelvin versus interlayer distance $d$ in nanometer for Fermi energy $V_g = 0.1\ \eV$.  From left to right the curves correspond to
$\tilde{U}/\tilde{U}_c = 0.5,0.7,0.9$.}
\end{figure}
\end{center}

As mentioned before, the use of a static screened Coulomb interaction is unreliable for
high energy intermediate states in the gap equation.  If we assume that our static screening estimate $U$ is
appropriate for $\ep < 2 V_{g}$ and another estimate $\tilde{U}$ is appropriate for $\ep > 2 V_{g}$, the
gap equation becomes
$$
\frac{1}{U} =\frac{\tilde{U}_c}{\tilde{U}_c-\tilde{U}} \,
\frac{1}{2} \int_{0}^{2V_{g}}  d \ep \nu(\ep) \sum_{s=\pm 1} \frac{1-2 n_f(\ep + s V_g) }{\ep + s V_g}.
$$
Here $\tilde{U}_c$ is the value of $\tilde{U}$ necessary to solve the $V_{g} = 0$ gap equation:
$$
\frac{1}{\tilde{U}_c} =  \nu(\xi),
$$
where $\nu(\xi)$ is the density of states evaluated at the ultraviolet cutoff. In Fig.~\ref{fig_phenpic} we show the $T_c$ versus $d$ lines for several values of $\tilde{U}/\tilde{U}_c$.
From left to right the curves correspond to $\tilde{U}/\tilde{U}_c = 0.5,0.7,0.9$.

\section{Discussion}
\label{sec_conc}
In this paper we have discussed a number of possible estimates for the critical temperature for
exciton condensation in Coulomb coupled graphene layers in which a gate voltage has
been applied to induce nesting between the conduction band Fermi level in the high-density
layer and the valence band Fermi surface of the low-density layer.  One of our main results concerns the
sublattice structure of the inter-layer coherence order, which tends to be mainly off-diagonal in
sublattice and opposite in sign for the $AA$ and $BB$ components.  This structure
is due to the momentum-direction dependence of inter-subband phases in both
conduction and valence band states near the $K$ and $K'$ valleys in graphene.
Similar observations have been made previously.\cite{seradjeh,Basu}
These extra phases cause contributions to the anomalous inter-layer
self-energy that are off-diagonal in sublattice to tend toward small values,
yielding approximate cancelation.  In a bilayer system this self-energy structure
leads to compatible self-energy contributions from coherence between the bands
with nested Fermi surfaces and from coherence between the two remote bands, the valence band of the high-density
layer and the conduction band of the low-density layer.  In addition the sublattice
structure of bands remains consistent in graphene out to energies well beyond the Fermi energies of the
nested bands, implying that contributions to interlayer coherence can be expected from high energy states
in both carrier and remote bands.

Critical temperature estimates for coherent graphene bilayers are considerably complicated by the
importance on the one-hand of screening by the gate-induced carriers at
low energies and on the other hand of lattice scale correlations at high energies.
For a two-dimensional system, mean-field theory overestimates the transition temperature to a superfluid and instead a KT-transition should be considered.  For strong short-range interaction this consideration
leads to a correction of the transition temperature by a factor $\sim 3$.\cite{lim}
We have reported on a number of $T_c$ estimates that are based on momentum and frequency
independent interactions.  When appropriate values for the effective interaction strength are
estimated based on low-energy screening considerations, large critical temperatures tend to
occur at low carrier densities simply because screening is then minimized.  These $T_c$ estimates likely
misrepresent trends as a function of carrier density, since screening at high energies is in fact
not strongly influences by carriers.  On the other hand, they do correctly,
capture the fact that spontaneous coherence would appear even in systems without carriers
if interactions were strong enough.  In the contact interaction approximation the gap equation
in the absence of carriers implies a Stoner-like criterion in which order appears for
$U \nu(\xi) >1$, where $\nu(\xi) \sim 1/(t a)^2$ is the density-of-states at the graphene Dirac model
ultraviolet cut-off. The interaction assumes this form because the density-of-states
in graphene increases linearly with energy over a very broad energy range.
It is interesting to compare this condition with the corresponding Stoner-like
condition for density-wave states in a single-layer graphene sheet, $U \nu(\xi) > 1$.\cite{singlest}

In the single-layer case, experiment seems to clearly indicate that the ground state is
not a density-wave state, {\em i.e.} that $U < \nu(\xi)^{-1}$.  On the other hand, the fact that
density-wave states appear to occur in the presence of a magnetic field may suggest that the
criterion is nearly met.  If so, Fig.~\ref{fig_phenpic} shows that gating-induced Fermi surface
nesting can enhance electron-hole pairing correlations and induce order at substantial temperatures.

The difference between our results and the prediction by Kharitonov {\it et al.}\cite{kharitonov,kharitonov2} that $T_c$ is maximally of the order of milliKelvins deserves further comment. Firstly, the value of the dielectric constant for the system we consider $\ep=1$ differs from the value $\ep \simeq $4 from Refs.~[\onlinecite{kharitonov}] and~\cite{kharitonov2}. This difference leads to a drastic difference in the prediction for $T_c$, as mentioned above. Furthermore, in their estimate of the upper limit for $T_c$, Kharitonov {\it et al.} set the width of the pairing region to zero, while we have shown that $T_c$ depends exponentially of the ratio of this width
to the Fermi energy. This distinction is important for the low Fermi energies we consider.   A similar effect leads to the enhancement of the gap at zero temperature, as shown in Ref.~[\onlinecite{lozovik2}].
None of these estimates include the effect of finite frequencies, which is likely to reduce screening effects and
raise transition temperatures further.

In conclusion, we have determined the sublattice structure of the condensate and
discussed the inter-layer coherence phase diagrams predicted by various estimates of interlayer
interaction strengths.  Our calculations make no assumptions about the sublattice structure of the condensate and take
the full graphene dispersion into account.  When contact interaction approximations are used high transition temperatures can occur.  For the separable-potential approximation, we find transition temperatures of the order of Kelvins for the parameter range $V_g=0.25-0.5\ \meV$ and $d<4\ \nm$. This results differs greatly from the result obtained by Kharitonov {\it et al.},\cite{kharitonov,kharitonov2} that $T_c$ is maximally of the order of milliKelvins, mainly because of the
important role played in our calculations by states at high energy intermediate states with $\ep > 2 V_{g}$.
Careful consideration of the roles of retardation effects and coherence in reducing screening could
increase $T_c$ estimates further.

\acknowledgments
This work was supported by the Stichting voor Fundamenteel Onderzoek der Materie
(FOM), the Netherlands Organization for Scientific Research (NWO), and by the European Research Council
(ERC). AHM was supported by DOE Division of Materials Sciences and Engineering grant
DE-FG03- 02ER45958 and by Welch Foundation
grant F-1672.\\

\section{Methods}
\label{sec_met}
In this section we describe the methods we used to obtain the results shown in Sec.~\ref{sec_res}. We use a functional-integral approach  since it provides a convenient starting point
to account for non-mean-field effects.

\subsection{Action of double-layer graphene}
The coherent state path-integral representation for the partition function $Z$ is given by
\be \label{eq_Z}
Z = \int d[\psi^*] d[\psi] e^{-(S_0[\psi^*,\psi] + S_I[\psi^*,\psi])/\hb},
\ee
where the integration is over all Grassmann-valued fields $\psi^*$ and $\psi$ that are anti-periodic
on the interval $[0,\hb \beta]$. The non-interacting action $S_0[\psi^*,\psi]$ in Eq.~(\ref{eq_Z}) is given by
\begin{multline} \label{eq_S0}
S_0[\psi^*,\psi]=
\sum_{\kv,\om_n} \sum_{\al,\al',\si,\si'}
\psi^*_{\al,\si}(\kv,i \om_n)\\
\times \left[-\hb G_{0;\al,\si;\al',\si'}^{-1} (\kv,i \om_n) \right] \psi_{\al',\si'}(\kv,i \om_n),
\end{multline}
where the momenta are restricted to the first Brillouin zone, the $\om_n = \pi(2 n+1)/\hb \beta$ are the fermionic Matsubara frequencies with $\beta = 1/k_B T$ the inverse temperature, $\al =A,B$ is the sublattice index, and $\si=t,b$ the which-layer pseudospin index. We ignore the electron spin for this moment, to return to it later. In the tight-binding model for graphene, the inverse noninteracting Green's function from Eq.~(\ref{eq_S0}) is
$$
G_{0;\si;\si'}^{-1}(\kv,i \om_n) = -\frac{\de_{\si,\si'}}{\hb}
\begin{pmatrix}
-i \hb \om_n -\mu_\si& f(\kv) \\
f^*(\kv)& -i \hb \om_n -\mu_\si
\end{pmatrix},
$$
where $\mu_\si$ is the chemical potential for the $\si$-layer and $f(\kv) = -t\left(1+e^{-i \kv \cdot \rv_1} + e^{-i \kv \cdot (\rv_1+\rv_2 )}\right)$, where $t = 2.8\ \eV$ the nearest-neighbor hopping strength, and $\rv_1$ and $\rv_2$ are the lattice vectors.\cite{neto,ARPES} The interaction contribution to the total action $S_I[\psi^*,\psi]$ in Eq.~(\ref{eq_Z}) describes the interaction between electrons via the Coulomb interaction and consists of both intra and interlayer terms. The effect of the intralayer terms is to renormalize the chemical potentials $\mu_\si$ and hopping strength $t$ and to screen the interlayer interaction. \cite{barlas,ARPES} These terms will be omitted in the remainder of this paper. To account for them, we assume that the renormalized chemical potentials are such that in the normal state the top layer is electron and the bottom layer hole-like, with equal densities. The corresponding Fermi levels are denoted by the carrier Fermi energy $V_g$. To incorporate the effect of screening by density fluctuations, we use the screened instead of the bare Coulomb interaction as the interlayer interaction. Then, $S_I[\psi^*,\psi]$ consists only of interlayer terms and is given by
\begin{multline} \label{eq_SI}
S_I[\psi^*,\psi] =\int_{0}^{\hb \beta} d \tau \sum_{\rv,\rv'} \sum_{\al,\al'}
V^{\scr}(\rv-\rv')\\
\times \psi_{\al,t}^*(\rv,\tau) \psi_{\al',b}^*(\rv',\tau)
\psi_{\al',b}(\rv',\tau) \psi_{\al,t}(\rv,\tau),
\end{multline}
where the position summations are over all $N$ unit cell positions. Since the transition depends only weakly on the precise stacking of the layers,\cite{zhang} we make the simplifying assumption that the stacking is such that the $A$-sites ($B$-sites) of the top layer lie directly above the $A$-sites ($B$-sites) of the bottom layer.

The statically screened interlayer Coulomb interaction is given by
\begin{align}\nonumber
V^{\scr}(\qv) &= \sum_{\rv} V^{\scr}(\rv) e^{i \qv \cdot \rv} \\ \label{eq_V}
&= \frac{V(q) e^{- q d} }{1 - 2 V(q) \Pi(q) +\left(1-e^{- 2 q d} \right) V^2(q) \Pi^2(q)},
\end{align}
where
$$
V(q)= \frac{e^2}{4 \pi \ep_0 \ep} \frac{1}{A} \frac{2 \pi}{q},
$$
is the bare interaction,\cite{lozovik} $\ep_0$ is the permittivity of the vacuum, $\ep$ is the dielectric constant of the surrounding medium which we take to be air with $\ep = 1$, $d$ is the distance between the graphene layers, and $A = 3 \sqrt{3} a^2/2$ the area of the graphene unit cell, with $a = 0.142 \ \nm$ the nearest-neighbor distance between the carbon atoms. The polarizability $\Pi$ is given in the Dirac approximation by $\Pi(q) = - 4 \nu(V_g) $ for $q<2 k_F$\\
which is the momentum range relevant for the carrier-band contribution to the gap equation.  The
factor 4 is due to spin and valley degeneracy, and the density of states is $\nu(V_g) = (A/2\pi) V_g/(3 a t/2)^2$. \cite{hwang} This form of $S_I[\psi^*,\psi]$ is valid in the long-wavelength approximation, where the interlayer Coulomb interaction is independent of the sublattice index. It will be convenient to introduce the dimensionless momentum variable $y = q d$, so that the dimensionless interaction $\ti{V}^{\scr}(y)$ becomes
\begin{align}\nonumber
\ti{V}^{\scr}(y) &\equiv \frac{1}{V(1/d)} V^{\scr}(y/d) \\ \label{eq_Vti}
&=\frac{e^{- y} }{y+ 2 \ga(\ep) k_F d+\left(1-e^{- 2 y} \right) (\ga(\ep) k_F d )^2/y},
\end{align}
where we defined
$$
-q V(q) \Pi(q) \equiv \ga(\ep) k_F = 9.66 k_F /\ep.
$$

\subsection{Derivation of the effective action}
Since the condensed state is a broken symmetry state, we cannot resort to perturbation theory to determine $T_c$. Instead, we perform a so-called Hubbard-Stratonovich transformation to obtain an effective action in terms of the order parameter for exciton condensation. Concretely, this procedure entails multiplying the partition function $Z$ from Eq.~(\ref{eq_Z}) by a Gaussian functional integral with value unity over the order parameter $\De_{\al,\al'}(\rv,\rv',\tau)$ that is on average given by $V^{\scr}(\rv-\rv') \la \psi_{t,\al}^*(\rv,\tau) \psi_{b,\al'}(\rv',\tau) \ra$. By an appropriate choice of the parameters in this integral, the interacting action $S_I[\psi^*,\psi]$ from Eq.~(\ref{eq_SI}) can be canceled from the argument of the exponent in the functional integral Eq.~(\ref{eq_Z}). Then, the integral over the electron fields can be performed analytically and an effective action is obtained in terms of the order parameter, which is given by
\begin{multline} \label{eq_Seff}
S_{\eff}[\De^*,\De] = - \hb \Tr \log \left(- \hat{G}^{-1}\right)\\
+\frac{\hb \beta}{N}
\sum_{\kv,\kv', \Kv,\om_{m}} \sum_{\al,\al'} \left(\frac{1}{V^{\scr}}\right)\left(\kv- \kv' \right) \\
\times \De^*_{\al,\al'}(\kv,\kv+\Kv,i \om_{m}) \De_{\al,\al'}(\kv',\kv'+\Kv,i \om_{m}),
\end{multline}
where $\om_m = 2 \pi m/\hb \beta$ are now the bosonic Matsubara frequencies, $\left(\frac{1}{V^{\scr}}\right)(\kv)$ is the Fourier transform of the inverse interaction in position space $1/V^{\scr}(\rv-\rv')$ and where
$\hat{G}^{-1} = \hat{G}^{-1}_0 - \hat{\Si}$ with the electron selfenergy given by
\begin{multline}
\hb \Si_{\al,\si;\al',\si'}(\kv,i \om_n;\kv',i \om_{n'}) = \\
 -\left[ \de_{\si,b} \de_{\si',t} \De_{\al',\al}(\kv',\kv,i\om_{n}-i \om_{n'})
\right. \\
\left.
+\de_{\si,t} \de_{\si',b} \De^*_{\al,\al'}(\kv,\kv',i \om_{n'}-i \om_{n})\right].
\end{multline}
We determine $T_c$ for a second-order phase transition to the condensed state by expanding the effective action $S_{\eff}[\De^*,\De]$ in Eq.~(\ref{eq_Seff}) to second order in the order parameter $\De^*$ and $\De$. Since we expect the order parameter to be translationally invariant in space and time, we only consider the contribution of the zero frequency and zero center-of-mass momentum components of the order parameter $\De^*$ and $\De$ to the effective action $S_{\eff}[\De^*,\De]$ in Eq.~(\ref{eq_Seff}). For notational simplicity, we define the shorthand $\De_{\al,\al'}(\kv,\kv,0) = \De_{\al,\al'}(\kv)$. This procedure yields
\begin{multline} \label{eq_Seff2}
S_{\eff}[\De^*,\De] = \hb \beta
\sum_{\kv,\kv'} \sum_{\al_1,\al_2,\al_3,\al_4}
\De^*_{\al_1,\al_2}(\kv)\\
\times M_{\al_1,\al_2,\al_3,\al_4}(\kv,\kv')
\De_{\al_3,\al_4}(\kv'),
\end{multline}
where
\begin{multline} \label{eq_M}
M_{\al_1,\al_2,\al_3,\al_4}(\kv,\kv') \\
=\frac{1}{N} \de_{\al_1,\al_3} \de_{\al_2,\al_4} \left(\frac{1}{V^{\scr}}\right)\left(\kv- \kv' \right) - \de_{\kv,\kv'} \Bv_{\al_1,\al_2,\al_3,\al_4}(\kv),
\end{multline}
with the interlayer polarization $\Bv$ given by
\begin{multline} \label{eq_B}
\Bv_{\al_1,\al_2,\al_3,\al_4}(\kv) \\
= - \frac{1}{\hb^2 \beta}
\sum_{\om_n} G_{0;\al_3,t;\al_1,t}(\kv,i \om_n) G_{0;\al_2,b;\al_4,b}(\kv,i \om_{n})\\
=\frac{1}{4}
\bpm
B_{0} & e^{i \phi} B_{1} & -e^{-i \phi} B_{1} & B_{2}\\
e^{-i \phi} B_{1} & B_{0} & e^{-2 i \phi} B_{2} & -e^{-i \phi} B_{1} \\
-e^{i \phi} B_{1} & e^{2 i \phi} B_{2} & B_{0} & e^{i \phi} B_{1}\\
B_{2} & -e^{i \phi} B_{1} & e^{- i \phi} B_{1} & B_{0}
\epm.
\end{multline}
Here we dropped the $\kv$ dependence of the $B_{i}$, defined $\phi = \arg[f(\kv)]$, and
\begin{align*}
B_{0} & = 2 B(+,+)\ +\ B(-,+)\ +\ B(+,-)\\
B_{1} & = B(-,+)\ -\ B(+,-)\\
B_{2} & = 2 B(+,+)\ -\ B(-,+)\ -\ B(+,-).
\end{align*}
Moreover
\be \label{eq_Bf}
B(s_t,s_b) = -\frac{n_f \biglb( \ep_{s_t,t}(\kv)\bigrb) -n_f\biglb(\ep_{s_b,b}(\kv)\bigrb)}{\ep_{s_t,t}(\kv)-\ep_{s_b,b}(\kv)},
\ee
with $s_t,s_b = \pm1$ and top and bottom layer dispersions
$$
\ep_{s,t}(\kv) = s |f(\kv)|+V_g
\qaq
\ep_{s,b}(\kv) = s |f(\kv)|-V_g,
$$
with $s= \pm1$. The letter $B$ is chosen in Eqs.~(\ref{eq_B},\ref{eq_Bf}) because Eq.~(\ref{eq_Bf}) is the expression of a bubble diagram, which describes screening by electron-hole pairs. We note that when we perform a particle-hole transformation in the (hole-like) top layer, the numerator of the right-hand side of Eq.~(\ref{eq_Bf}) becomes $1-n_f(\ep_{s_1,t}(\kv)) -n_f(\ep_{s_2,b}(\kv))$, and we obtain the familiar expression for the ladder diagram from BCS theory.

Now, we comment on the importance of the real electron spin. Each independent fermion species contributes to the screening of the Coulomb interaction. Therefore, a factor of two due to the spin degeneracy should be included in the expression for the polarizability $\Pi(q)$, as we did above. The interlayer Coulomb interaction is to a very good approximation independent of spin, and we need to consider how the effective action Eq.~(\ref{eq_Seff2}) changes if we include the electron spin, in Eq.~(\ref{eq_S0}) and Eq.~(\ref{eq_SI}). The electron spin can be incorporated in our formalism by extending the definition of $\al$ to include both the sublattice and spin quantum numbers, so that the order parameter has 16 components. The noninteracting Greens functions in the expression for $\Bv$ in Eq.~(\ref{eq_B}) are diagonal in spin, so that the same is true for $\Bv$ in Eq.~(\ref{eq_B}) and $M$ in Eq.~(\ref{eq_M}). Thus, we find that the contributions to the effective action of the 4 spin pairing channels are decoupled and thus we find four identical $T_c$ equations, one for each channel. Thus, it is correct for the determination of the phase diagram to ignore the electron spin in our formalism.

\subsection{Derivation of the linearized gap equation}
The transition temperature $T_c$ is now given by the maximum temperature for which the matrix $M$ defined in Eq.~(\ref{eq_M}) has a zero eigenvalue, or equivalently the maximal temperature for which for all $\kv$, $\al_1$, and $\al_2$ we have
$$
\sum_{\kv'} \sum_{\al_3,\al_4} M_{\al_1,\al_2,\al_3,\al_4}(\kv,\kv') \De_{\al_3,\al_4}(\kv')=0.
$$
To get rid of the Fourier transform of the reciprocal interaction, we multiply with $V^{\scr}(\kv''-\kv)$ and sum over $\kv$ to obtain the gap equation
\begin{multline} \label{eq_gap}
\De_{\al_1,\al_2}(\kv)=
\frac{1}{N} \sum_{\kv',\al_3,\al_4} V^{\scr}(\kv-\kv') \\
\times
\Bv_{\al_1,\al_2,\al_3,\al_4}(\kv') \De_{\al_3,\al_4}(\kv').
\end{multline}
Below, we use two methods to find approximate solutions of Eq.~(\ref{eq_gap}), namely by modeling the interaction as a contact interaction and using the separable-potential approximation.\\

We remark that the screened interaction in Eq.~(\ref{eq_gap}) should in first instance be evaluated at frequency
$\omega = \epsilon_{\kv}-\epsilon_{\kv'}$. The frequency and wavevector arguments which appear in the remote band part of this integral are ones for which our static screening approximation is not reliable. It is, however, not immediately clear how to improve on the approximation we employ because of corrections to the simple RPA screening function and the role of $\sigma$ and $\sigma^*$ bands that we do not consider. We therefore choose to use the
static RPA screening functions in this paper, but
remain cognizant of limitations in the predictive power of our (or any other) semi-analytic $T_c$ calculation.

\subsection{Approximation 1: Contact interaction}
The contact-interaction approximation is rather crude for the Coulomb interaction and can only be used to obtain qualitative results for the phase diagram and the condensate structure. In this approximation, we replace the interaction matrix elements in momentum space $V^{\scr}(\kv-\kv')$ in Eq.~(\ref{eq_gap}) by an effective strength $U$ which is an appropriate average of $V^{\scr}$ over its arguments.  It then follows that the components of the order parameter $\De_{\al_1,\al_2}(\kv)$ are independent of momentum. This approach was used previously for this system .\cite{seradjeh,lozovik2}  However, because the contact interaction averages out the structure of the Coulomb interaction, one may only expect to obtain qualitative results using this approximation. Setting $V^{\scr}(\kv-\kv')=U$ in Eq.~(\ref{eq_gap}), we obtain a $4 \times 4$ matrix equation
$$
\De_{\al_1,\al_2}=
\frac{U}{N} \sum_{\kv',\al_3,\al_4}
\Bv_{\al_1,\al_2,\al_3,\al_4}(\kv') \De_{\al_3,\al_4},
$$
so that the critical condition is that the $4\times4$ matrix $\bm{\Pi}=(U/N) \sum_{\kv}\Bv(\kv)$ has eigenvalue 1. We have that
$$
\bm{\Pi} = \frac{U}{4N}
\bpm
\Pi_{0,0} & \Pi_{1,1} & -\Pi_{1,1} & \Pi_{2,0}\\
\Pi_{1,1} & \Pi_{0,0} & \Pi_{2,2} & -\Pi_{1,1} \\
-\Pi_{1,1} & \Pi_{2,2} & \Pi_{0,0} & \Pi_{1,1}\\
\Pi_{2,0} & -\Pi_{1,1} & \Pi_{1,1} & \Pi_{0,0}
\epm,
$$
where we defined
$$
\Pi_{i,l} = \frac{1}{N} \sum_{\kv} B_{i}(\kv) \cos\{l \arg[f(\kv)]\}.
$$
The eigenvalues of $\Pi$ can be computed in closed form.  Setting the largest eigenvalue $\to 1$
yields the following $T_c$ equation:
\begin{multline} \label{eq_Tc1}
1 = \frac{U}{8}\left(2 \Pi_{0,0}-\Pi_{2,0}-\Pi_{2,2} \phantom{\sqrt{\Pi^2}} \right.\\
\left.+\sqrt{16 \Pi^2_{1,1}+(\Pi_{2,0}-\Pi_{2,2})^2}\right).
\end{multline}
We remark that the eigenvector corresponding to this largest eigenvalue has opposite $(A,A)$ and $(B,B)$ components, a result previously found in a model without intersublattice components for the order parameter $\De_{A,B}$ and $\De_{B,A}$. \cite{seradjeh}

\subsubsection{Dirac approximation}
It is interesting to consider the result for the critical temperature Eq.~(\ref{eq_Tc1}) in the Dirac approximation and compare its solution to the critical temperature obtained using the full dispersion. The linear Dirac spectrum is often used as an approximation to the dispersion of graphene, where one sets $f(\kv) = (3 a t/2)(k_x+ik_y)$. Then, the $B_{i}(\kv)$ depend only on the length of $\kv$ and it follows that the $\Pi_{i,l}$ vanish for nonzero angular momentum $l$. Then, Eq.~(\ref{eq_Tc1}) can be written as
\begin{align} \label{eq_Tc1D}
1 &= \frac{U}{2 N} \sum_{\kv} \left[B(-,+)\ +\ B(+,-) \right] \nonumber  \\
&=\frac{U}{2} \int_{0}^\xi d \ep \nu(\ep) \sum_{s=\pm 1} \frac{1-2 n_f(\ep + s V_g) }{\ep + s V_g},
\end{align}
where we used that $n_f(-\ep) = 1 - n_f(\ep)$, a factor 2 was added in the second line to account for the presence of the valley degeneracy in the momentum integral, and $\xi$ is some high-energy cutoff, on which we comment below. Again, $\nu(\ep)$ is the density of states for a single valley and spin species. The equation Eq.~(\ref{eq_Tc1D}) describes the situation of band-diagonal pairing, as described previously,\cite{lozovik2} in which there is no pairing between the close-lying conduction bands, and far-laying valance bands. Following an approach used to analyze superconductivity in single-layer graphene, \cite{kopnin} we can evaluate Eq.~(\ref{eq_Tc1D}) further and obtain
\begin{multline} \label{eq_Tc1D2}
\frac{y}{\lam}+x \int_{y-x}^{y+x} d x' \frac{\tanh (x')}{x'} - \log(\cosh(y-x)\cosh(y+x))  \\
=2 x \int_{0}^{x} d x' \frac{\tanh (x')}{x'}-
2 \log\cosh(x),
\end{multline}
with $ x = \beta V_g/2$, $y = \beta \xi/2$ and the dimensionless coupling constant given by $\lam = U \nu(\xi)/2$. The system has a quantum critical point only for $V_g=0$ and $U_{\QCP} = 6.25\ \eV$ where $\lam = 1/2$ and $U_{\QCP} \nu(\xi) = 1$, which is a Stoner criterion for the spontaneous polarization of the valence bands in the two layers which are filled for $V_g=0$. An important point is that the solution of Eq.~(\ref{eq_Tc1D2}) depends on the high- energy cutoff $\xi$, and only when we choose a particular value for $\xi$, are we able to compare the results obtained using the Dirac approximation and the full dispersion. We find this value of $\xi$ by demanding that the interaction strength $U_{\QCP}$ at which the quantum critical point occurs for the Dirac approximation in Eq.~(\ref{eq_Tc1D2}) coincides with the value of $U_{\QCP}$ obtained from Eq.~(\ref{eq_Tc1}). This equality leads to the equation
$$
\nu(\xi) = \frac{1}{2N} \sum_{\kv} \frac{1}{|f(\kv)|},
$$
which yields $\xi = 6.83\ \eV$. The results obtained using this procedure are discussed in Sec.~\ref{sec_res} and shown in Fig.~\ref{fig_pdUT1}. When $y \gg x \gg 1$ we may approximate Eq.~(\ref{eq_Tc1D2}) to obtain a BCS-like result for the transition temperature
\be \label{eq_tcexp}
k_b T_c= \frac{V_g}{2} \frac{\sqrt{\xi-V_g}}{\sqrt{\xi+V_g}} \exp\left(
- \frac{1}{U \nu(V_g)}+ \frac{\xi}{V_g} +C-1 \right),
\ee
where $C = \lim_{R\to \infty}\left\{ \int_{0}^R [\tanh(x)/x] d x - \log(R) \right\} = 0.82.
$ Apart from the usual BCS term in the exponent $-1/U \nu(V_g)$, we also find an additional term which scales as the length of the pairing region $\xi$ over the Fermi energy $V_g$. This effect will also influence the $T_c$ found in our separable-potential approximation and one of the reasons that we predict a higher value of $T_c$ for small Fermi energies as compared to Refs.~[\onlinecite{kharitonov}]~and~[\onlinecite{kharitonov2}]. We finally note that the gap equation on the close-band approximation can be obtained from Eq.~(\ref{eq_Tc1D}) by only taking the $s=-1$ term of the summation. The close-band results are also discussed discussed in Sec.~\ref{sec_res} and shown in Fig.~\ref{fig_pdUT1}. We note that an equation similar to Eq.~(\ref{eq_tcexp}) was found in Ref.~[\onlinecite{lozovik2}] for the magnitude of the gap at zero temperature.

\subsubsection{Estimation of $U$}
By estimating the effective interaction strength $U$ as a function of the interlayer distance $d$, we may transform the horizontal axis in Fig.~\ref{fig_pdUT2} and obtain the phase diagram with the transition temperature versus $d$. We estimate $U$ by evaluating the angular average of the screened interaction Eq.~(\ref{eq_V}) over the incoming and outgoing momenta restricted to the Fermi surface, $\kv$ and $\kv'$, respectively
\begin{align}
U(d) &= \frac{\int_{\ep_\kv,\ep_{\kv'} = V_g} d \kv d \kv' V^{\scr}(\kv-\kv') }{\int_{\ep_\kv,\ep_{\kv'} = V_g} d \kv d \kv' }\\
& =\frac{1}{\pi} \int_0^{ \pi} d \phi V^{\scr}\biglb(2 k_F \sin(\phi)\bigrb),
\end{align}
which can be easily evaluated numerically.

\subsection{Approximation 2: Separable approximation}
\begin{center}
\begin{figure}
\includegraphics[width = 0.47 \textwidth]{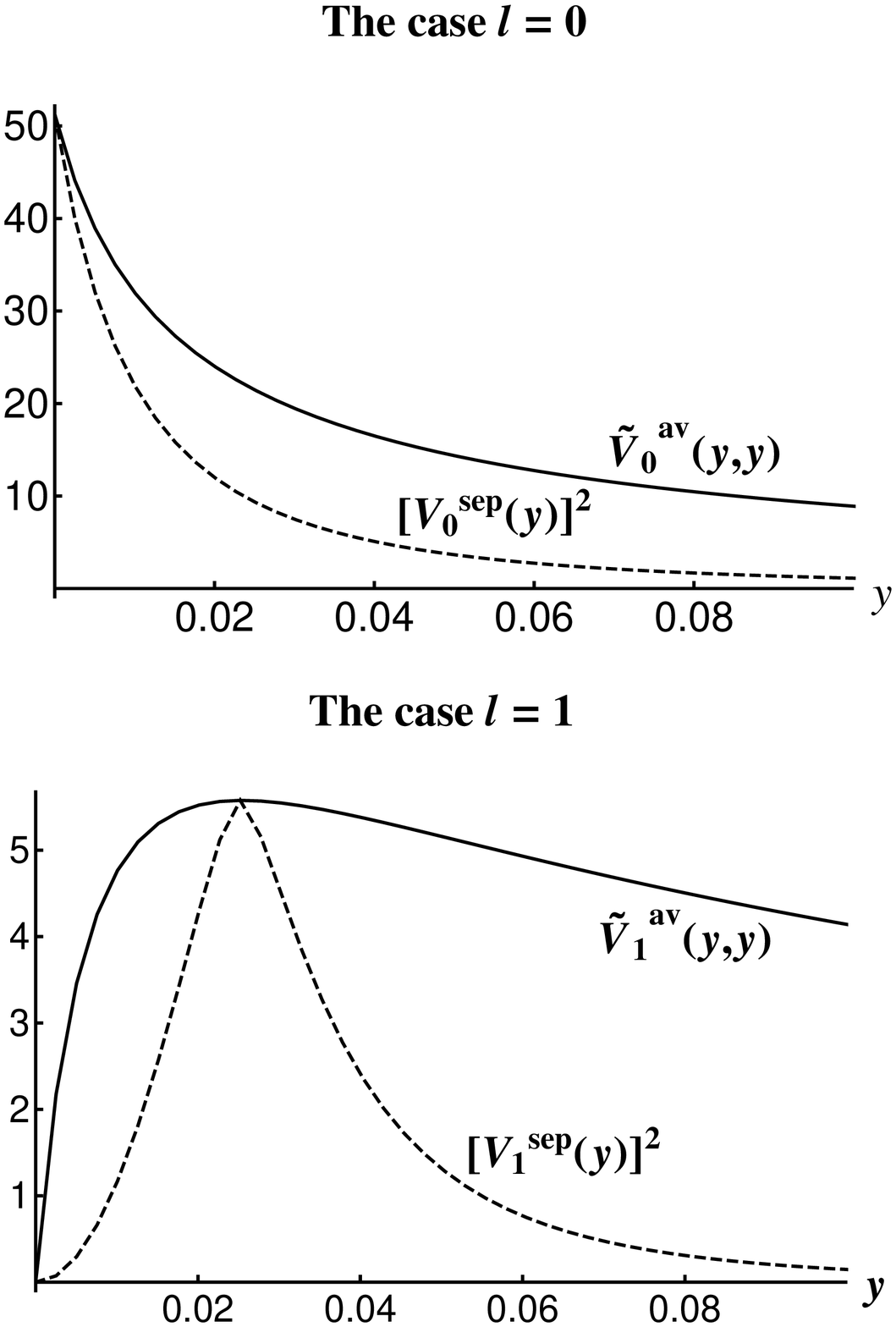}
\caption{\label{fig_sepap}
We plot $\ti{V}^{\avv}_{l}(y,y)$ and $\left[V^{\sep}_{l}(y)\right]^2$ by the solid and dashed line, respectively, for the cases $l=0$ (top graph) and $l=1$ (bottom graph).}
\end{figure}
\end{center}

In order to obtain a quantitative prediction for the mean-field transition temperature, it is not sufficient to approximate the screened Coulomb interaction by a contact interaction. Instead, we will approximate the screened Coulomb interaction by a function which is separable in the incoming and outgoing momenta, as was also done in Ref.~[\onlinecite{lozovik2}] to determine the gap at zero temperature. We show below how to implement this procedure concretely. We will consider $s$-wave solutions for the gap functions of the form
\be \label{eq_Dan}
\De_{\al_1,\al_2}(y)= \De_{\al_1,\al_2}(y) e^{i l_{\al_1,\al_2} \phi },
\ee
where $\phi$ is the azimuthal angle of $\kv$, the $l_{\al_1,\al_2}$ is the angular momentum quantum number, and we transformed to the dimensionless momenta $y=kd$. Since we showed that the influence of the full dispersion is small, we continue in the Dirac approximation, where we may rewrite the gap equation Eq.~(\ref{eq_gap}) as
\begin{multline} \label{eq_sgap}
\De_{\al_1,\al_2}(y) =
\frac{A}{2 \pi d^2} \sum_{\al_3,\al_4} \int y' d y' V^{\avv}_{\al_1,\al_2,\al_3,\al_4}(y,y') \\
\times \Bv^R_{\al_1,\al_2,\al_3,\al_4}(y') \De_{\al_3,\al_4}(y'),
\end{multline}
where we defined the angular averaged interaction
\begin{multline}\label{eq_Vavv}
V^{\avv}_{\al_1,\al_2,\al_3,\al_4}(y,y') =
\frac{1}{2 \pi} \int d \phi'
e^{i(l_{\al_3,\al_4}\phi'-l_{\al_1,\al_2} \phi)}\\
\times e^{i n_{\al_1,\al_2,\al_3,\al_4} \phi'} V^{\scr}\biglb(\sqrt{y^2 +y'^2 -2 \cos(\phi'-\phi)y y'}\biglb),
\end{multline}
where we anticipated the fact that the $l_{\al,\al'}$ will be chosen such that the right hand side of Eq.~(\ref{eq_Vavv}) does not depend on $\phi$. We furthermore defined the radial part of $\Bv$ from Eq.~(\ref{eq_B}) as $\Bv^R$ in the following way
$$
\Bv_{\al_1,\al_2,\al_3,\al_4}(\yv/d) = \Bv^R_{\al_1,\al_2,\al_3,\al_4}(y) e^{i n_{\al_1,\al_2,\al_3,\al_4} \phi},
$$
where $\Bv$ should be considered in the Dirac approximation and the values of the integers $n_{\al_1,\al_2,\al_3,\al_4}$ can be read of from the expression for $\Bv$ in Eq.~(\ref{eq_B}). Note that we did not include an extra factor 2 due to the valley degeneracy, since we assume that there is no intervalley scattering. Then, we may argue that each valley pairing channel leads to an equivalent gap equation, and we may ignore the valley label, similar as we did above for the electron spin. The right-hand side of Eq.~(\ref{eq_sgap}) should be independent of $\phi$, which is the case when
$$
l_{A,A} = l_{B,B} = l_0, \quad
l_{A,B} = l_0 -1, \qaq l_{B,A} = l_0+1.
$$
for some integer $l_0$. Then, the gap equation Eq.~(\ref{eq_sgap}) can be written in a simplified form as
\begin{multline} \label{eq_s2gap}
\De_{\al_1,\al_2}(y) =
\frac{\ga(\ep)}{4} \sum_{\al_3,\al_4} \int y' d y' \ti{V}^{\avv}_{l_{\al_1,\al_2}}(y,y') \\
\times \ti{\Bv}_{\al_1,\al_2,\al_3,\al_4}(y') \De_{\al_3,\al_4}(y'),
\end{multline}
where
\begin{multline*}
\ti{V}^{\avv}_{l}(y,y') = \frac{1}{2 \pi} \int d \phi \cos(l \phi) \\
\times \ti{V}^{\scr}\biglb(\sqrt{y^2 +y'^2 -2 \cos(\phi)y y'}\bigrb),
\end{multline*}
where $\ti{V}^{\scr}(y)$ was defined above in Eq.~(\ref{eq_Vti}). We defined $\ti{\Bv}^R$ as dimensionless form of $\Bv^R$, i.e. with the $B(s_1,s_2)$ from Eq.~(\ref{eq_Bf}) replaced by the dimensionless $\ti{B}$ defined by
$$
\ti{B}(s_t,s_b) = - \frac{ \frac{1}{1+\exp[\beta' (s_t y - k_F d)]} - \frac{1}{1+\exp[\beta' (s_b y + k_F d)]}}{(s_t y-k_F d)-(s_b y + k_F d)},
$$
with $\beta'= \beta \hb v_f/d$. From these expressions we obtain that for a {\it fixed} value of $k_F d$ the transition temperature goes as $1/d$ and increases linearly with the Fermi momentum $k_F$ and the carrier Fermi energy $V_g$. Up to this point, our rewriting of the gap equation Eq.~(\ref{eq_gap}) is exact, under the ansatz Eq.~(\ref{eq_Dan}). In order to be able to obtain numerical results, we now approximate $\ti{V}^{\avv}_{l}(y,y')$ by a function that is separable in $y$ and $y'$. This approximation amounts to choosing a function $V^{\sep}$ such that
\be \label{eq_sepan}
\ti{V}^{\avv}_{l}(y,y') \simeq V^{\sep}_{l}(y) V^{\sep}_{l}(y').
\ee
From Eq.~(\ref{eq_s2gap}) we see that the $y$ dependence of $\De_{\al_1,\al_2}(y)$ in this case goes as $V^{\sep}_{l_{\al_1,\al_2}}(y)$ so that it natural to define $\De_{\al_1,\al_2}(y) = V^{\sep}_{l_{\al_1,\al_2}}(y) \De'_{\al_1,\al_2}$. The gap equation Eq.~(\ref{eq_s2gap}) thus becomes a $4\times4$ matrix equation independent of $y$
\begin{multline} \label{eq_s3gap}
\De'_{\al_1,\al_2} =
\frac{\ga(\ep)}{4} \sum_{\al_3,\al_4} \int y' d y' V^{\sep}_{l_{\al_1,\al_2}}(y') V^{\sep}_{l_{\al_3,\al_4}}(y') \\
\times \ti{\Bv}_{\al_1,\al_2,\al_3,\al_4}(y') \De'_{\al_3,\al_4}.
\end{multline}
After choosing a functional form of $V^{\sep}_{l}$ we can find the transition temperature as the largest temperature for which Eq.~(\ref{eq_s3gap}) has a solution. For our purposes, it is sufficient to choose the following form of $V^{\sep}_{l}$
$$
V^{\sep}_{l}(y) = \frac{V^{\avv}_{l}\left(y,y^{\reff}_{l}\right)}{\sqrt{V^{\avv}_{l}\left(y^{\reff}_{l},y^{\reff}_{l}\right)}},
$$
where $y^{\reff}_{l}$ is some reference momentum. It is natural to choose $y^{\reff}_{l}$ as the position of the maximum of $V^{\avv}_{l}(y,y)$, which yields
$$
y^{\reff}_{l=0} = 0 \qaq
y^{\reff}_{l=1} \simeq 2.61 \ga(\ep) k_F d.
$$
To gain insight in the effect of this approximation, we plot in Fig.~\ref{fig_sepap} the functions $\ti{V}^{\avv}_{l}(y,y)$ and $\left[V^{\sep}_{l}(y)\right]^2$, which would fall on top of each other if the approximation Eq.~(\ref{eq_sepan}) were exact. In Fig.~\ref{fig_sepap} we plot $\ti{V}^{\avv}_{l}(y,y)$ and $\left[V^{\sep}_{l}(y)\right]^2$ by the solid and dashed line, respectively, for the cases $l=0$ (top graph) and $l=1$ (bottom graph). Since $\left[V^{\sep}_{l}(y)\right]^2$ is always lower than $\ti{V}^{\avv}_{l}(y,y)$ one expects that the transition temperatures found in our analysis are a lower boundary for the mean-field transition temperature for exciton condensation.

\end{document}